\documentclass[aps,prl,superscriptaddress,twocolumn,showpacs]{revtex4}
\usepackage{graphicx}

\newcommand{\be}{\begin{equation}}
\newcommand{\ee}{\end{equation}}
\newcommand{\bea}{\begin{eqnarray}}
\newcommand{\eea}{\end{eqnarray}} 
\newcommand{\dxx}{\partial_{X\hbox{\hskip-.1cm}X}}

\begin{document}

\title{When does coarsening occur in the dynamics of one-dimensional fronts ?}

\author{Paolo Politi}
\email{politi@ifac.cnr.it}
\affiliation{Istituto di Fisica Applicata ``N. Carrara",
Consiglio Nazionale delle Ricerche, Via Madonna del Piano,
50019 Sesto Fiorentino, Italy}
\affiliation{Istituto Nazionale per la Fisica della Materia,
UdR Firenze, Via G. Sansone 1, 50019 Sesto Fiorentino, Italy} 
\affiliation{Laboratoire de Spectrom\'etrie Physique-GREPHE, CNRS,
UJF-Grenoble 1, BP87, F-38402 Saint Martin d'H\`eres, France}

\author{Chaouqi Misbah}
\email{chaouqi.misbah@ujf-grenoble.fr}
\affiliation{Laboratoire de Spectrom\'etrie Physique-GREPHE, CNRS,
UJF-Grenoble 1, BP87, F-38402 Saint Martin d'H\`eres, France}

\date{\today}

\begin{abstract}
Dynamics of a one-dimensional growing front  with an unstable
straight profile are analyzed. We argue that a coarsening process occurs 
if and only if the period $\lambda$ of the steady state solution 
is an increasing function of its amplitude $A$. 
This statement is rigorously proved for two important classes of 
conserved and nonconserved models by investigating 
the phase diffusion equation of the steady pattern. 
We further provide clear numerical evidences 
for the growth equation of a stepped crystal surface.
\end{abstract}




\maketitle  

{\it Introduction -}
One of the most fascinating feature of nonequilibrium
systems is their ability to build up a spatial (generally) ordered pattern from
an initially structurless state when some control parameter reaches
a critical value. Typical examples can be found in hydrodynamics,
crystal growth,
the Turing reaction-diffusion systems, and so on \cite{cross93}. 
These pattern
forming systems can be broadly classified into two important categories:
(i) the first category includes
the systems which select a typical pattern length scale $\lambda$ which
is fixed  for  a given value of the  control parameters. In that
case the whole pattern needs not to be steady in time
but can undergo oscillations, or even exhibit chaotic dynamics (a prominent example
is the Kuramoto-Sivashinsky dynamics \cite{KS}). What matters
is the selection of a length scale.
(ii) The second
category corresponds to the situation where 
there is  a perpetual coarsening, in that the wavelength $\lambda(t)$
increases without bound in the course of time. 
Interrupted coarsening may also happen for special situations (see below).
One must exclude the trivial case where the bifurcation occurs at a wavenumber
$q=q_c$ and the band $\Delta q$ of active modes extends from $q_c-\delta_1$ 
to $q_c+\delta_2$. If $\Delta q=\delta_1 +\delta_2$ is not too large, 
the system selects, of course, a pattern with a given intrinsic lengthscale. 
Examples include Turing patterns, Rayleigh-B\'enard convection...,
close enough to the instability threshold. Our discussion focuses on 
situations where there is no lower boundary on the active wavenumbers 
(arbitrarily long wavelengths are active). It is only
at very large distance from the instability threshold that, for example, 
the Turing and Rayleigh-B\'enard systems exhibit nontrivial dynamics 
(chaos, coarsening of turbulent plumes...), and this is
precisely due to the fact that higher and higher harmonics become active.

A central question in nonequilibrium
pattern forming systems is to identify
general criteria (if any) for the presence of coarsening or not
without resorting to a systematic time-dependent calculation.
We put forward a criterion   for some
classes of nonlinear equations which is uniquely based on the
behavior of the steady-state solutions. 
This criterion is derived on the basis of the analysis
of the phase diffusion equation \cite{cross93},
$\partial_T\psi = D\dxx\psi$,
$\psi$ being the phase.
A negative sign of the phase diffusion coefficient, $D$, is therefore
a signature for the systems that undergo coarsening~\cite{note_Dnegative}.
The second and more important step will be
to relate the sign of 
$D$ with the derivative, $\partial\lambda/\partial A$, of the period
$\lambda$ of the stationary solution with respect to its amplitude $A$.
We will show that the sign of $D$ is opposite to that  of
$\partial\lambda/\partial A$. A coarsening process can therefore take
place, $D<0$, if and only if $\partial\lambda/\partial A$ is positive.
This relation will be proved analytically for two classes of nonconserved and
conserved equations (Eqs.~(\ref{GL}) and (\ref{CH}) respectively) 
and checked numerically for an equation which can display 
different dynamical scenarios with varying some physical parameters
(Eq.~(\ref{eq:FG})).
Finally, considerations based on the Lyapunov functional will complement our
analysis.

{\it Nonconserved equations -}
We begin with  the following class
of nonconserved equations~\cite{note_noise}:
\be
\partial_t u = \partial_x^2u - F(u) \equiv L[u] ~ ,
\label{GL}
\ee
where $\partial_t,\partial_x$ indicate the time and the spatial
derivatives, respectively.
$F(u)$ is a generic function, playing the role of a force.
We require that $F(u)\approx -u$ at small $u$, so the trivial
solution $u(x,t)=0$  is linearly unstable 
against fluctuations of wavelength $\lambda>\lambda_0=2\pi$.
The  Ginzburg-Landau (GL) equation is recovered for  $F(u)=-u+u^3$, so
Eq.~(\ref{GL}) will be referred to as the generalized GL equation.

We can now remark that stationary solutions, $u_0(x)$, satisfy the condition
\be
\partial_x^2u_0 - F(u_0) =0 ~ ,
\label{GL_sta}
\ee
and $u_0(x)$ corresponds to the motion of a particle in the potential
$V(u)=-\int du F(u)$. For a linearly unstable profile, $V(u)\approx 
u^2/2$ at small $u$.

The steady-state $u_0(x)$ is a periodic solution with 
the spatial periodicity $\lambda$, 
$u_0(x+\lambda)=u_0(x)$. It is useful to consider
the phase of the pattern, $\phi$, and we shall
use a nonlinear WKB-like analysis~\cite{cross93} for the derivation
of the phase equation. For a steady-state periodic solution,
$\phi=qx$ where $q=2\pi/\lambda$. In general the pattern
will evolve in time, and thus the phase $\phi(x,t)$ is a function
of space and time. We consider long wavelength modulations (which
are the most dangerous ones owing to translational invariance
of the initial pattern), and let $\epsilon$ denote the smallness
of the phase modulation. We introduce
a slow phase $\psi$ which is a function of a slow spatial
variable $X\equiv \epsilon x$. Because we expect any deviation
from the periodic pattern to evolve diffusively we anticipate that  the slow
time $T$ scales as $T=\epsilon^2 t$. The slow phase $\psi(X,T)$ is 
related to the fast phase $\phi(x,t)$ by 
$\phi=\psi/\epsilon$
so that  the local wavenumber is given
by
$q(X,T)={\partial \phi/ \partial x}={\partial \psi / \partial X}$.
In a multiscale spirit we have to make the substitutions
\begin{equation}
{\partial_x \rightarrow q\partial_\phi +\epsilon \partial_X } \; ,\;\;\;
{\partial_t \rightarrow \epsilon^2 \partial_T \phi \partial_\phi +
\epsilon^2 \partial_T} ~ .
\end{equation}
We then expand $u=u_0+\epsilon u_1 +....$ in powers of $\epsilon$. To leading order
$\epsilon^0$ we find that $u_0$ obeys $L[u_0]=0$. For the generalized
GL equation we have 
\begin{equation}
q^2 \partial_{\phi\phi}u_0 - F(u_0) =0 ~ .
\label{GLs}
\end{equation} 

The next order reads $\tilde L[u_1]=f(u_0)$, where $\tilde L$ is the
(Fr\'echet) derivative~\cite{Zwillinger} 
of $L$ and $f$ is the inhomogeneous term arising
from phase modulation. For the generalized GL equation,
$\tilde L= q^2 \partial_{\phi\phi} -F'(u_0)$, and
\begin{equation}
f(u_0) =\partial _T \psi\partial_\phi 
u_0- [\partial_\phi u_0 +2q\partial_{\phi  q}u_0]\dxx \psi ~ .
\end{equation}    

It is easily shown that, because of translational invariance, 
the linear operator $\tilde L$ has the nontrivial
solution $\partial_\phi u_0$. It follows 
that $u_1$ exists only if the inhomogeneous term $f(u_0)$ is orthogonal 
to $\partial_\phi u_0$~\cite{Fredholm}. 
This condition results in the sought after
diffusion equation
\begin{equation}
\partial_T\psi =D\dxx\psi ~ ,
\end{equation}
where $D$, the phase diffusion coefficient, is given by
\begin{equation}
D =
{ \partial_q\langle q(\partial_\phi u_0)^2\rangle \over 
\langle(\partial_\phi u_0)^2\rangle }
\equiv {D_1\over D_2} ~ .
\label{D_GL}
\end{equation}
Here above, $\langle \dots\rangle =(2\pi)^{-1}\int_0^{2\pi} ... d\phi$ 
is the inner product,
the denominator $D_2$ is always positive and the sign of $D$ is fixed
by the numerator $D_1=\partial_q\langle q(\partial_\phi u_0)^2\rangle$. 

Eq.~(\ref{GLs}), which is equivalent to Eq.~(\ref{GL_sta}),
corresponds to the equation of motion of a particle of unitary mass,
subject to the force $F(u_0)$: $u_0$ is the spatial coordinate and
$\tau=\phi/q$ is the time. So, we have:
\be
\langle q(\partial_\phi u_0)^2\rangle = 
{1\over 2\pi}\int_0^{2\pi/q} 
\hbox{\hskip -.3cm}
d\tau (\partial_\tau u_0)^2 =
{J\over 2\pi} ~ ,
\ee
where we have introduced the action variable $J$.

Finally, if we remember \cite{Goldstein} 
that the period of the oscillatory motion
is given by the derivative of the action $J$ with respect to the energy
$(\lambda=\partial J/\partial E)$, we find:
\be
D_1 = {1\over 2\pi} {\partial J\over\partial q} = 
-{\lambda^3\over 4\pi^2}
\left({\partial\lambda\over\partial E}\right)^{-1}  ~ ,
\ee
so that the sign of $D_1$ is opposite to that  of
${\partial \lambda/\partial E}$: there is coarsening $(D< 0)$
if and only if the wavelength is an increasing function of the
energy of the particle or, equivalently, of the amplitude $A$, which is 
related to the energy by $E=V(A)$.

{\it Conserved equations -}
We now apply the same procedure to a class of conserved equations,
which---in the same spirit---will be called generalized
Cahn-Hilliard (CH) equations:
\be
\partial_t u = -\partial_x^2 [\partial_x^2 u - F(u) ] \equiv
-\partial_x^2 L[u] ~ .
\label{CH}
\ee

The linear stability analysis of the trivial solution $u=0$ gives the spectrum
$\omega=q^2-q^4$ and allows to define the fastest growing mode, 
$\lambda_c=\sqrt{2}\lambda_0$.
We skip the calculations to attain the phase equation 
and just give the final expression
for the phase diffusion coefficient, 
\be
D = {q^2\partial_q\langle q(\partial_\phi u_0)^2\rangle \over
\langle u_0^2\rangle } \equiv {q^2 D_1\over \tilde D_2} ~ .
\label{D_CH}
\ee

The $q^2$ factor at the numerator is distinctly due to the
conservation law, i.e. to the operator $-\partial_x^2$ in front
of $L$ in Eq.~(\ref{CH}).
The sign of $D$ is fixed by the quantity $D_1=
\partial_q\langle q(\partial_\phi u_0)^2\rangle$ as for the nonconserved case.
Our criterion is therefore proved for the generalized CH equation as well.

Similarly to $J=\oint d\tau (\partial_\tau u_0)^2$ we can define 
$I = \oint u_0^2(\tau)$, where $\oint d\tau$ means---in the mechanical
analogy---the time integral on an oscillation period.
Using these quantities, 
$D_2 = \langle (\partial_\phi u_0)^2\rangle = \lambda J/ 4\pi^2$ and
$\tilde D_2 = \langle u_0^2\rangle = I/\lambda$.
Finally, the diffusion coefficient $D$ reads
\bea
D = - {\lambda^2\over J(\partial_E\lambda) } && ~~~
\hbox{\tt nonconserved (GL) models ~~} \\
D = - {\lambda^2\over I(\partial_E\lambda) } && ~~~
\hbox{\tt conserved (CH) models}  
\eea

The expressions clearly show that $D$ is negative if and only if 
$\partial_E\lambda > 0$.

{\it The different scenarios -}
Here above we have established for two important
classes of evolution equations that
the phase diffusion equation is stable (no coarsening) or unstable
(coarsening), according to the sign of $d\lambda/dA$.
Therefore, the following scenarios can be advanced
(see Fig.~1).
(i)~If $\lambda(A)$ is a decreasing function the system typically develops
a profile whose characteristic wavelength
keeps constant in the course of time.
This happens, e.g., if $F=-u-u^3$, i.e. $V(u) = u^2/2 + u^4/4$.
(ii)~If $\lambda(A)$ is an increasing function the system exhibits coarsening.
This is what happens in the standard GL and CH models: $F=-u+u^3$.
(iii)~If $\lambda(A)$ starts to increase and afterwards it decreases the
system is expected to display 
coarsening that stops at the maximum of $\lambda(A)$.
This behavior is midway between (i) and (ii)
and it is called interrupted coarsening.
(iv)~If $\lambda(A)$ starts to decrease and afterwards it increases the 
system essentially behaves as in (ii).

\begin{figure}
\includegraphics[width=7cm,clip]{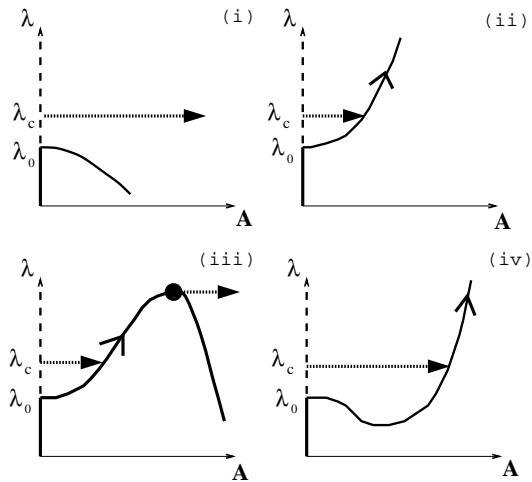}
\caption{Schematic wavelength-amplitude diagrams for the stationary
solutions of period $\lambda$ and amplitude $A$.
The four scenarios refer to different behaviors for $\lambda'(A)$,
as explained in the main text. The linear stability analysis
of the trivial solution $u=0$ is represented by the line $A=0$:
stability (full line) for $\lambda<\lambda_0=2\pi$ and instability
(dashed line) for $\lambda >\lambda_0$. 
For the conserved models the flat interface
starts to develop a profile of period $\lambda_c=\sqrt{2}\lambda_0$
(see the big dotted arrows close to the $\lambda$-axis).
In cases (ii)-(iv) a coarsening process takes place, but in case
(iii) it stops when $\lambda(A)$ attains the maximum
(interrupted coarsening) and the amplitude starts growing
(see the big dotted arrow close to the maximum).}
\end{figure}

It is now interesting to consider a growth equation of physical interest, 
which does not fall into the classes we have discussed (see Ref.
\cite{Review} for a general reference on crystal growth dynamics).
The equation has the general form
$\partial_t z = - \partial_x j$,

\be
\partial_t z(x,t) = -\partial_x \left\{
B(m) + G(m)\partial_x [C(m)\partial_x m] \right\} ~ ,
\label{eq:FG}
\ee
where $m=\partial_x z$ is the local slope of the front and
plays the role of $u$ in Eq.~(\ref{CH}). It is worthnoting that 
Eq.~(\ref{eq:FG}) reduces to Eq.~(\ref{CH}) if $C(m)$ and $G(m)$
are constant.
Equations of this kind are often met in nature, such as
in epitaxial growth of high symmetry as well as vicinal 
surfaces \cite{Review,Gillet00}. 
They are
also encountered in the study of sand ripples \cite{Csahok00}, in 
problems of spinodal decomposition, in the study of many 
linear defects, and so on. 

It is possible to define the new variable 
$M(m) = \int^m \hbox{\hskip-.1cm}ds\, C(s)$, so that
the stationarity condition
$j=0$ can be rewritten as $B(M)+G(M)\partial_{xx}M=0$. Steady states therefore
correspond to the solution of the
Newton's equation for a fictitious particle, moving in the potential
\be
V(M) = \int\hbox{\hskip-.1cm} dM\, {B(M)\over G(M)} ~ .
\ee

Here we are interested in the growth dynamics of stepped, vicinal crystal
surfaces~\cite{note_vicinal} by vapor phase deposition techniques.
With reference to Eq.~(\ref{eq:FG}), $m$ is the
local slope of the step and \cite{thesis_FG}:
\bea
&& B(m) = {m\over 1+m^2} \; , ~
G(m) = {1+\beta\sqrt{1+m^2} \over (1+\beta)(1+m^2)}~ , \\
&& C(m) = {1+c(1+m^2)(1+2m^2)\over (1+c)(1+m^2)^{3/2}} ~ , 
\eea
where $\beta$ and $c$ are positive, adimensional parameters.
The former, $\beta$, is the relative strength between line diffusion
and terrace diffusion, and the latter, $c$, is the strength of the 
elastic coupling between steps.
The linear spectrum reads $\omega=q^2 -q^4$.
We now report the results of our analysis of $\lambda(A)$ for 
the different values of the parameters $\beta$ and $c$ (see also
Fig.~1).

If elastic interactions are absent, $c=0$, $\lambda$ is always a
decreasing function of $A$ (regime (i)).
If elastic interactions are weak, $0<c < 1/(1+2\beta)$, $\lambda(A)$
decreases at small slopes and increases at large slopes (regime (iv)).
For strong elastic interactions, $c> 1/(1+2\beta)$, $\lambda(A)$ is
always an increasing function (regime (ii)).

 From a dynamical point of view, we therefore expect no
coarsening for $c=0$ and perpetual coarsening for $c\ne 0$.
The case without elastic interactions has been considered in
Ref.~\cite{OPL} for $\beta=0$ and in Ref.~\cite{Gillet00}
for any $\beta$: Refs.~\cite{OPL,Gillet00} 
clearly show the absence of coarsening and the
development of a constant pattern length scale with $\lambda=\lambda_c$.
The case $c\ne0$ has been treated in Ref.~\cite{Paulin} (see Figs.~1,2
therein) and a persistent coarsening is revealed,
both for $c<1/(1+2\beta)$ (regime (iv)) and for $c>1/(1+2\beta)$
(regime (ii)).

Recently, Eq.~(\ref{eq:FG}) has been modified in order to take into
account anisotropy effects \cite{Danker03}. In this case, $\lambda(A)$
may present a maximum, which leads to regime (iii).
As expected by our criterion, the coarsening is found~\cite{Danker03} 
to interrupt at the  maximum of $\lambda (A)$,
where the amplitude  grows unstably without bound.

An interrupted coarsening process was also found in a 
discrete model describing the growth dynamics of a
one-dimensional surface, the so-called Zeno model~\cite{Zeno}.
The surface is described as a sequence of steps which move 
following the Burton, Cabrera and Frank dynamics~\cite{BCF}.
What is relevant in our context is that the
interruption of  coarsening at  $\lambda=\lambda^*$ is related to the 
disappearance of stationary configurations 
for $\lambda>\lambda^*$~\cite{Zeno} (Fig.~1, scenario (iii)).

{\it Lyapunov functional -}
It is well known~\cite{Chaikin} that the two classes of nonconserved,
Eq.~(\ref{GL}), and conserved, Eq.~(\ref{CH}), models 
can be derived by the Lyapunov functional~\cite{note_V}
${\cal F}[u(x,t)] = \int dx \left[ {1\over 2} (\partial_x u)^2 - V(u)\right]$,
through the relations
$\partial_t u = -(\delta {\cal F} / \delta u)$ (GL models)
and $\partial_t u = \partial_x^2 (\delta {\cal F} / \delta u)$
(CH models).
It is also easy to check that $d{\cal F}/dt\le 0$ in both cases, i.e. dynamics
proceeds so as to minimize ${\cal F}$. 

It is trivial that $d{\cal F}/dt$ vanishes for the stationary
configurations of period $\lambda$, $d_t {\cal F}[u_\lambda(x)]=0$.
Rather, we are interested here to study the dependence of ${\cal F}$ on the
period $\lambda$ of the steady state. 
The relation ${\cal F}[u_\lambda(x)]/\ell = (J/\lambda -E)$ is 
found, where $\ell$ is the length of the growing front, and
the derivative is easily evaluated~\cite{note_der},
$\ell^{-1}(d {\cal F}[u_\lambda(x)] /d \lambda) = -(J/\lambda^2) < 0 $.
This result, together with the relation $d{\cal F}/dt\le 0$ which ensures
that dynamics minimizes ${\cal F}$, complements the scenarios described
in Fig.~1: if $\lambda$ increases with $A$ there is coarsening,
and if $\lambda(A)$ has a maximum at $\lambda=\lambda^*$ coarsening
stops at $\lambda^*$.

{\it Perspectives -}
Our analysis has focused on equations where either coarsening takes place, 
or the wavelength is frozen while the amplitude `diverges' depending 
on whether $\lambda (A)$ is an increasing or decreasing function.
This does not exhaust  all possible  scenarios that occur in nonequilibrium
systems. A prominent example is the Kuramoto-Sivashinsky (KS) equation
which is known to  produce spatio-temporal chaos with 
a dominant length scale (corresponding to the fastest growing mode).
It would therefore be desirable to see how the steady-branch $\lambda (A)$ 
would behave in the presence of KS-dominated dynamics.
We have done a preliminary study~\cite{long} of the modified CH equation
\cite{Golovin}:
$\partial_t u = -\partial_x^2 [\partial_x^2 u +u-u^3] 
+\nu  u \partial _x u$. 
For $\nu=0$ this equation reduces to the  CH one, while for large $\nu$ 
an appropriate rescaling leads to the KS one.
Above a critical value $\nu_0$ the branch $\lambda (A)$ undergoes a fold 
singularity where $\lambda (A)$ exhibits a  turning point.
We have found that in correspondence of that
an anticoarsening process appears, that is to say $\lambda(t)$ decreases. 
Our work is in progress
and it aims to clarify possible relations between the sign of
$\lambda'(A)$ and the type of dynamics exhibited by the system.

A more general puzzling question is whether, for extended systems enjoying basic symmetries
(translation in the plane, $x\rightarrow x+x_0$ and $u\rightarrow u+u_0$
(or $z\rightarrow z+z_0$), and parity)
coarsening, chaos  or `diverging' amplitudes
with a frozen periodicity are the only possible scenarios. 
Is there any simple link between symmetries and the kind of dynamics
(order or disorder)
that a system exhibits when it does not succumb coarsening?

A final question concerns the diffusion coefficient $D$ when
(perpetual) coarsening occurs, $D<0$. In this case, the
typical pattern length scale $\lambda$ increases in time. Is it possible
to derive the coarsening law, $\lambda(t)$, from the knowledge
of $|D(\lambda)|$?
It will be an important task to clarify all the above questions
and to consider the possible extension of our arguments 
to higher dimension as well as to nonlocal equations,
as they arise for example in solidification and viscous fingering.

We gratefully acknowledge useful discussions with several
members of the Florentine Dynamics of Complex Systems group.


\begin{thebibliography}{99}
\bibitem{note_Dnegative}
In principle a negative $D$ could entail a decreasing of $\lambda$ (splitting)
as well. However, this is inconsistent with the result $d{\cal F}/d\lambda <0$
(see the Section on the Lyapunov functional) and with the stability of the
flat interface at small lengthscales.
\bibitem{note_noise}
Noise does not modify the coarsening scenarios discussed throughout the paper.
\bibitem{note_vicinal}
A vicinal surface is a surface which is slightly miscut from
a high symmetry orientation. It looks like a flight of stairs
with steps of atomic height separating large terraces.
\bibitem{note_V}
${\cal F}$ is usually written in terms of $U(u)=-V(u)$, but 
here is more useful to use $V(u)$.
\end{thebibliography}
\end{document}